\documentclass[showpacs,amsmath,amssymb,aps,showkeys,floatfix,prd,a4paper]{revtex4}
\usepackage{subfigure}
\usepackage{dcolumn}
\usepackage{bm}
\usepackage{epsfig}
\usepackage{amsfonts}
\usepackage{amssymb,amscd}
\def\lsim{\raise0.3ex\hbox{$<$\kern-0.75em\raise-1.1ex\hbox{$\sim$}}}
\def\gsim{\raise0.3ex\hbox{$>$\kern-0.75em\raise-1.1ex\hbox{$\sim$}}}

\newcommand{\be}{\begin{equation}}
\newcommand{\ee}{\end{equation}}
\newcommand{\ba}{\begin{eqnarray}}
\newcommand{\ea}{\end{eqnarray}}

\def\beq{\begin{equation}}
\def\eeq{\end{equation}}
\def\beqa{\begin{eqnarray}}
\def\eeqa{\end{eqnarray}}

\def\gappeq{\mathrel{\rlap {\raise.5ex\hbox{$>$}}
{\lower.5ex\hbox{$\sim$}}}}
\def\lappeq{\mathrel{\rlap{\raise.5ex\hbox{$<$}}
{\lower.5ex\hbox{$\sim$}}}}
\def\Toprel#1\over#2{\mathrel{\mathop{#2}\limits^{#1}}}

\begin{document}
\begin{flushright}
\vskip1cm
\end{flushright}

\title{Two and three photon fusion into  charmonium in  ultra-peripheral nuclear collisions}

\author{R. Fariello$^{1,4}$, D. Bhandari$^{2}$, C.A. Bertulani$^{2,3}$,   and  F.S. Navarra$^{2,4}$\\ 
$^{1}$ Departamento de Ci\^encias da Computa\c{c}\~ao, 
Universidade Estadual de Montes Claros, Avenida Rui Braga, sn, 
Vila Mauric\'eia, CEP 39401-089, Montes Claros, MG, Brazil.\\
$^2$Department of Physics and Astronomy, Texas A\&M University-Commerce, 
Commerce, Texas 75429, USA\\
$^3$Institut f\"ur Kernphysik,  Technische Universit\"at Darmstadt, 64289 Darmstadt, Germany\\
$^4$Instituto de F\'{\i}sica, Universidade de S\~{a}o Paulo, 
Rua do Mat\~ao 1371 - CEP 05508-090, 
Cidade Universit\'aria, S\~{a}o Paulo, SP, Brazil\\
}

\begin{abstract}
In this paper we investigate the production of charmonium states in 
two and three photon fusion processes in nucleus -- nucleus  
collisions at the CERN Large Hadron Collider (LHC) energies. Our results 
indicate that the experimental study of these processes is feasible and can  
be used  to constrain the theoretical decay widths and give information on 
the non $c - \bar{c}$ components of these states.  
\end{abstract}



\date{\today}

\maketitle

\section{Introduction} 

During the last twenty years dozens of new charmonium states have been 
observed at the LHC \cite{nora20,nos19,zhu19,meiss18,espo17,hosa16,nn14}.
Some of them are, beyond any doubt, multiquark (or exotic) states, i.e.,   
states in which the minimum quark content is $c \bar{c} q \bar{q}$. This 
is the case of all charged exotic states \cite{nn14}. Among the charge neutral 
states there are some which are, for several reasons, incompatible with the 
$c \bar{c}$ configuration. This is the case of the most famous exotic state, 
the $X(3872)$, which is now called $\chi_{c1}(3872)$. There are other charge 
neutral states, whose multiquark nature is still under debate, such as the 
$\psi(3770)$.  

The central discussion in this field is about the internal structure of the 
multiquark 
states.  The most often studied  configurations are 
the  meson molecule  and the tetraquark. The main difference between a 
tetraquark and  a meson molecule  is that the former is compact and the 
interaction between the constituents occurs through color exchange  forces 
whereas the latter is an extended object and the interaction between its 
constituents  happens through meson exchange forces 
\cite{nora20,nos19,zhu19,meiss18,espo17,hosa16,nn14}.

One aspect that is sometimes forgotten, is that, being quantum systems, these 
states can be mixtures. There may be  charmonium-tetraquark, 
charmonium-molecule or tetraquark-molecule mixtures. Here again, different 
works which consider these multiquarks states as mixtures do not reach a   
consensus. For example, in the case of the well studied $\chi_{c1}(3872)$, 
in Ref. \cite{zanetti} the mass and strong decay 
width were very well  reproduced assuming that it has a $c \bar{c}$ 
component with a weight of  97 \% and a tetraquark component with     
$3$ \% weight. On the other hand, in Ref. \cite{chines} it was 
shown that, in the case of production in proton-proton collisions, 
the best description of the data could be achieved with a 
charmonium-molecule combination, i.e. $\chi_{c1}' - D \bar{D}^*$, 
in which the $c \bar{c}$ component is of the order of $28 - 44$ \%. In spite  
of the discrepancies, it is remarkable that in both works a large  
$c \bar{c}$ component is required to explain data. 

The study of exotic states started in $B$ factories and then went to 
hadron colliders. The hadronic production of exotic states became a new 
way to discriminate between different configurations. The production of  
$\chi_{c1}(3872)$ in proton-proton collisions in the pure molecular approach 
was studied in \cite{espo2,espo3,grin18}. In \cite{espo4}, the analysis of 
recent data from the LHCb with the comover interaction model favored the 
compact tetraquark configuration. An attempt to use the pure tetraquark 
model to study $\chi_{c1}(3872)$ and $T_{4c}$ ($X(6900)$) production in 
proton-proton collisions was  presented in \cite{tetradps}.  All these works  
have improved our understanding of these new states, but there are still 
important questions to be answered. 

The very recent publication of the CMS collaboration \cite{cms22}
reporting the observation of the $\chi_{c1}(3872)$ in Pb-Pb collisions opened 
a new era in the study of exotics in heavy ion collisions. The main advantage 
of using heavy ion projectiles is the very large  number of produced          
$c - \bar{c}$ pairs. In the case of central collisions, the main disadvantage 
is that the total number of produced particles is very large and it becomes 
difficult to search for the multiquark states. 

It is also possible to study multiquark states in ultra-peripheral collisions  
(UPCs). High energy hadrons are an intense source of photons (For a review see 
Ref. \cite{BB88,BB94,BKN05,Bal08,hencken,Baur02}). At large impact parameters 
($b > R_{h_1} + R_{h_2}$), the photon -- induced interactions become dominant  
with the final state being  characterized by the multiquark state  and the  
presence of two intact hadrons if the resonance was produced in two or 
three-photon interactions. Experimental results at the LHC               
\cite{alice, alice2,lhcb,lhcb2} have shown that the study of photon --     
induced interactions in hadronic collisions is feasible and can be used to 
improve our understanding of the QCD dynamics. The idea of studying exotic   
meson production in UPCs was pioneered in \cite{bertu}, where the production 
cross section of several light and heavy well known mesons  
in nucleus-nucleus collisions was computed. Later, in Refs. \cite{vicwer}     
and \cite{mbgn16}, the same formalism was applied to the production of mesons 
and heavy exotic states in pp, pA and AA collisions.


In this work we will focus on  $c - \bar{c}$ states, giving special 
attention to the states, which are presently quoted by the PDG \cite{pdg} 
as $c - \bar{c}$, but whose nature is still under debate and which might still 
be multiquark states, or at least, might have a multiquark 
(either tetraquark or molecular) component. We will argue that we can use 
photon fusion processes in UPCs to confirm (or not) their $c - \bar{c}$ 
nature. This is possible because in these processes we only use QED and a well
established method to project quark-antiquark pairs into bound states, avoiding
some model dependence inherent to hadronic processes. We will revisit and 
update the calculations performed in \cite{bn02} and include new states. 
We will study the most recently observed particles using the quantum number 
assignments published in the most recent compilation made by the 
Particle Data Group \cite{pdg}. We shall consider both  two-photon and 
tree-photon processes.  As it will be seen, all the ingredients of the 
calculation are fixed. The formalism developed in \cite{bn02} applies to 
fermion-antifermion systems, being thus appropriate to the study of conventional 
quarkonium states. We will also apply it to the controversial cases, where     
the multiquark nature of the state is still under debate. A future experimental 
confirmation of our predictins would establish the quark-antiquark nature of 
these states. 

This paper is organized as follows. In section II we present a short  
description of the formalism used for particle production in 
two-photon interactions at hadronic colliders. In section III 
we discuss meson production  in three-photon interactions. In all  cases
we present the update of the results obtained for the production of 
charmonium in Pb-Pb collisions, including new states and making predictions 
for  LHC. Finally, in section IV we present a brief summary and  discussion 
of the results.

\section{Two photon processes}

The theoretical treatment of UPCs in relativistic heavy ion collisions has 
been extensively discussed  in the literature 
\cite{BB88,BB94,BKN05,Bal08,hencken,Baur02}. In what follows we will only review 
the main formulas needed to make predictions for  meson production 
in two and three photon interactions, which were derived in \cite{bn02}.

The differential cross section for the production of C-even mesons through 
two-photon fusion is given by \cite{bn02}:
\be
{d\sigma \over d P_z} = {16 (2J+1)\over \pi^2} {Z^4 \alpha^2 \over M^3}
\ {\Gamma_{\gamma\gamma} \over E}
\int d{\bf q}_{1t} d{\bf q}_{2t} \
({\bf q}_{1t}\times {\bf q}_{2t})^2 \,\,  
{\left[ F_1(q_{1t}^2)F_2(q_{2t}^2)\right]^2 \over
\left(q_{1t}^2+\omega_1^2/\gamma^2\right)^2
\left(q_{2t}^2+\omega_2^2/\gamma^2\right)^2}
\label{meson2}
\ee
where $P_z$, $E$, $M$ and $J$ are the longitudinal momentum, energy, mass and 
spin of the produced meson, respectively; $\Gamma_{\gamma\gamma}$ is 
the two-photon decay width of the meson; $Z$, $\alpha$ and $\gamma$ are the 
atomic number, the fine structure constant and the Lorentz 
factor. Finally, $F_1$ and $F_2$ are the projectile and target form factors. 
Following \cite{bn02} it is easy to relate the meson variables with the photon
energies $\omega_1$ and $\omega_2$:
\be
E = \omega_1 + \omega_2 \ , \ \ \ \
\omega_1 - \omega_2 = P_z \ , \ \ \ \ \
{\rm and} \ \ \ \ \
\omega_1 \omega_2 = M^2/4
\nonumber
\ee
The photon energies $\omega_1$ and $\omega_2$  are related to the mass
$M$ and the rapidity $Y $ of the outgoing  meson by
$\omega_1 = \frac{M}{2} e^Y$ and $\omega_2 = \frac{M}{2} e^{-Y}$.

As it was already mentioned, $F(q^2)$ is the nuclear form factor and it plays 
a crucial role in this formalism. The precise form of the form factor is the 
main source of uncertainties in our calculations. The Woods-Saxon distribution,  
with central density $\rho_0$, size $R$ and diffuseness $a$ gives a good 
description of the  densities of the nuclei.  Fortunately, this distribution 
is very well described by the convolution of a hard sphere and an Yukawa function
 \cite{nix}. In this case, the form factors can be calculated analytically:
\be
F(q^2) =
{4 \, \pi \, \rho_0 \over A \, q^3} \
\left[ \sin (qR) - qR \cos (qR) \right]
\ \left[ {1\over 1+q^2a^2}\right]
\label{formfact} .
\ee
For Pb we use $R = 6.63$ fm  and $a = 0.549$ fm, with $\rho_0$
normalized so that $\int d^3 r \rho (r) = 208$ \cite{Ja74}.
With the above expressions it is easy to compute the total cross sections 
with an adequate form factor \cite{nix}. 

During the derivation of the above formula for the cross section, we had to
use a prescription to bind together the produced quark and antiquark into a
bound state. We did this using  the projection operators \cite{bn02}
\ba
&\bar{u} \cdots  v \longrightarrow
\ \displaystyle{\Psi(0)\over 2\sqrt{M}}
\ {\rm tr} \left[\cdots (\not\!\! P + M) i \gamma^5\right] \nonumber \\
& \bar{u} \cdots  v \longrightarrow
\ \displaystyle{\Psi(0)\over 2\sqrt{M}}
\ {\rm tr} \left[\cdots (\not\!\! P + M) i \not\!\hat{e}^* \right]
\label{traces}
\ea
where $\cdots$ denotes any matrix operator. The first equation applies to spin 0 
and the second to spin 1 particles, respectively. In these equations 
$\Psi({\bf r})$ is the bound state wavefunction calculated at the origin  and 
$\hat{e}^*$ is the polarization vector of the outgoing vector meson.
Squaring the corresponding amplitude yields the factor $|\Psi(0)|^2$, which is 
then related to the decay width $\Gamma_{\gamma \gamma}$ through the formula 
derived by Van Royen and Weisskopf in Ref.~\cite{RW67} 
(see the discussion in \cite{bn02}) for fermion-antifermion annihilation. 
Hence, because of the hadronization         
prescription, the cross section formulas derived in \cite{bn02} apply to 
quark-antiquark states. Nevertheless, in order to obtain a first estimate 
we shall use the Van Royen - Weisskopf formula also for states, which are 
very likely  multiquark states, such as the $X(6900)$.

In what follows we will compute the production cross sections for conventional  
$c - \bar{c}$ and also to states whose status is still under debate. Therefore 
our results will serve as baseline for the experimental search in UPCs. If our 
predictions are confirmed this will be an argument in favor of the 
quark-antiquark assignment. If there are large discrepancies between data and 
our numbers, this will indicate the existence of a molecular or tetraquark 
component. As mentioned in the introduction, charge neutral states can always be
mixtures and in the existing calculations involving mixtures, 
the $c - \bar{c}$ component is always large. Hence our calculations will be 
relevant. Our strategy is conservative. Instead of creating a model for the 
production of multiquark states, we stick to the well know QED amplitudes 
complemented by experimental information. 

In Table \ref{tab1-23} we show the  cross sections for the production of $C$ 
even mesons in Pb-Pb collisions at $\sqrt{s_{NN}}= 5.02$ TeV 
using the formalism described above. Comparing with the results obtained in 
\cite{bn02} we observe 
some discrepancies, which are due to the use of updated values of the measured 
decay widths. We have also included the results for the $\eta_c(2S)$, which 
was not so well measured at that time.  

\begin{table}[h!]
\begin{center}
\begin{tabular}{|l|c|c|c|}    
\hline
State & Mass & $\Gamma_{\gamma\gamma}$[keV] & $\sigma ^{\text{LHC}}$[mb] 
\tabularnewline
    \hline
    \hline
     $\pi_0$ , $0^{-+}$ & 134 & 0.0078 &   38.0  
\tabularnewline
    $\eta$ , $0^{-+}$   & 547 & 0.46   &   17.3   
\tabularnewline
    $\eta '$ , $0^{-+}$ & 958 & 4.2    &   21.8   
\tabularnewline
    f$_2$ ,  $2^{++}$  & 1275 & 2.4    &   22.4   
\tabularnewline
   a$_2$ , $2^{++}$    & 1318 & 1.0    &    8.3    
\tabularnewline
$\eta_c$ , $0^{-+}$    & 2984 & 5.15   &    0.43   
\tabularnewline
$\chi_{0 c}$ , $0^{++}$ & 3415 & 2.2   &    0.11    
\tabularnewline
$\chi_{2 c}$ , $2^{++}$ & 3556 & 0.56  &    0.02   
\tabularnewline
$\eta_c (2S)$ , $0^{-+}$ & 3637 & 2.14 &    0.09
\tabularnewline
\hline
\end{tabular}
\caption{Cross sections for production of  C-even mesons in Pb-Pb 
ultra-peripheral collisions at $\sqrt{s_{NN}}= 5.02$ TeV. 
The decay widths are taken from the PDG \cite{pdg}.} 
\label{tab1-23}
\end{center}
\end{table}

In Table \ref{tab2-23} we show the update of the results obtained in 
\cite{mbgn16}  for the production cross section of the $J=0$ and $J=2$   
particles. In the PDG compilation, the quantum numbers and the 
nature of the $X(3940)$ are  still undefined and its two-photon decay width 
was not measured. We have used the theoretical values calculated in       
\cite{branz,oset}. The states $\chi_{c0}(3915)$ and $\chi_{c2}(3930)$ are 
treated as conventional
$c - \bar{c}$ scalar and tensor states respectively. However these assignemnts 
have been questioned (see for example, \cite{simo23}). In  Table \ref{tab2-23} 
we included results for the very recently measured $X(6900)$ state \cite{6900}. 
This state was seen in the $J/\psi$-$J/\psi$ invariant mass 
spectrum and therefore it could be a $c \bar{c} c\bar{c}$ state. After
the observation there was a series of works discussing its structure and  
hadronic production. Among them, Ref.~\cite{gm6900} is of special relevance to 
us. The authors have used the equivalent photon approximation (EPA) and the 
convolution formula included in the Appendix for the sake of discussion. The
formalism described there is quite similar to the one described above and the 
use of the Low formula for the $\gamma \gamma \to R$ reaction is equivalent to
using the Van Royen - Weisskopf formula. Since in Ref.~\cite{gm6900} the 
authors did not know the two-photon decay width of the $X(6900)$, they could 
not be very precise in their estimate. 
Later, this information was extracted from the analysis of light-by-light 
scattering in Ref.~\cite{pasca22}. The main observation was that the fit of the
measured $\gamma \gamma$ invariant mass spectrum becomes much better when one 
adds a resonance of mass $\simeq 6900$ MeV. Using different assumptions in the 
analysis they arrive at the values of $\Gamma_{\gamma \gamma}$  quoted in 
Table~\ref{tab2-23}, where I stands for interference and NI for no-interference
(for more details see \cite{pasca22}).  These values are surprisingly large an 
when inserted into our formulas yield  very large production cross sections.

\begin{table}[h!]
\begin{center}
\begin{tabular}{|c|c|c|c|c|c|}
\hline
State & Mass & $\Gamma_{\gamma\gamma}$[keV] & 
\multicolumn{3}{c|}{$\sigma$[$\mu$b]} \tabularnewline
\cline{4-6}
 & & &  $2.76$ TeV & $5.02$ TeV & $39$ TeV \tabularnewline
    \hline
    \hline
    X(3940), 0$^{++}$ & 3943 & 0.33 \cite{branz,oset}  &  5.5 &  9.7 &  32.5  \tabularnewline
    X(3940), 2$^{++}$ & 3943 & 0.27  \cite{branz,oset} & 22.5 & 39.6 & 133.0  \tabularnewline 
$\chi_{c2}(3930)$ , 2$^{++}$ & 3922 & 0.08 \cite{pdg} &  7.1 & 12.4 &  41.7  \tabularnewline
$\chi_{c0}(3915)$ , 0$^{++}$ & 3919 & 0.20 \cite{pdg}  &  3.4 &  6.0 &  20.1  \tabularnewline
    X(6900), $\,\,$(I)  & 6900 & 67 \cite{pasca22}  &   120.5  &  238 & 912    \tabularnewline
    X(6900), (NI) & 6900 & 45 \cite{pasca22}  & 81    &  160 &  612  \tabularnewline
   \hline
\end{tabular}
\caption{Cross sections for production of C-even  charmonium states  in  
Pb-Pb ultra-peripheral collisions at different energies. The highest energy 
might be relevant for collisions at the FCC \cite{fcc}.}
\label{tab2-23}
\end{center}
\end{table}

\eject

\section{Three-photon processes} 

The formalism described in the previous section is analogous to the 
equivalent photon approach  and the cross section could be rewritten 
as the well known convolution  EPA formula (given in the Appendix) for 
the process $\gamma \gamma \to R$, where $R$ is any integer spin resonance.
However, this formula can only be used for the production of $J = 0$ or $2$ 
states. For the case of vector meson production we need the  three-photon 
fusion process. In Ref. \cite{bn02} we derived the expression for the cross
section of three-photon fusion into a C-odd meson. In differential form it 
reads:
\ba
{d\sigma \over d P_z} &=&
1024 \, \pi \,  \Big|\Psi(0)\Big|^2
(Z\alpha)^6 { 1\over M^3E}
\
\int  { dq_{1t} \ q_{1t}^3\ \left[F(q_{1t}^2)\right]^2 \over
\left(q_{1t}^2 + \omega_2^2/\gamma^2 \right)^2}\nonumber \\
&\times&
\ \int  {dq_{2t}\ q_{2t} \
\left[F(q_{2t}^2)\right]^2 \over
 \left[ q_{2t}^2+(2\omega_1-\omega_2)^2/\gamma^2\right]^2} \,\,
\left[ \int {dk_t \ k_t\ F(k_t^2) \over
\left(k_t^2+(\omega_1-\omega_2)^2/\gamma^2\right)}
\right]^2
\ea
The definition of the variables are as in (\ref{meson2}). However, 
in the present case, the wave function $|\Psi(0)|^2$ can no longer 
be related to the $\gamma\gamma$ decay width. On the other hand, vector  
mesons can decay into $e^+e^-$ pairs and these decay widths are very well 
known experimentally. Following a similar derivation as for the 
$\gamma\gamma$ decay, 
the $e^+e^-$ decay width of the vector mesons can be shown \cite{RW67}  
to be proportional to the wave function squared, i.e. 
$\Gamma_{e^+e^-} \propto|\Psi(0)|^2$. Using the relation derived in 
\cite{RW67} we arrive at \cite{bn02}:
\be
\sigma =  \int d\omega \,  96\, \pi \, {\Gamma_{e^+e^-}\over M^3} \, 
                           {n(\omega)\over \omega}\ \ H (M,\omega)
\ee
where  $n(\omega)$ is given by:
\be                                                                               
n_(\omega) =  {2\over \pi}\ Z^2 \alpha \ \int {dq \ q^3                      
\ \left[ F(q^2) \right]^2 \over                                                 
\left(q^2+\omega^2/\gamma^2\right)^2} \ .                                       
\label{epaf2}                                                                     
\ee
and 
\be
H(M,\omega)  = Z^4\alpha^3 M^2
\int
{dq_{2t}\ q_{2t} \ \left[F(q_{2t}^2)\right]^2
\over
\left[ q_{2t}^2+(M^2/2\omega-\omega)^2/\gamma^2\right]^2} \,\,
\left[ \int {dk_t \ k_t\ F(k_t^2) \over
k_t^2+(M^2/4\omega-\omega)^2/\gamma^2}\right]^2
\label{three5}
\ee

In Table \ref{tab4-23} we present the cross sections for vector charmonium 
production. The first four lines are just an update of the results found in 
\cite{bn02}. The other lines present states which may be exotic. A common 
feature shared by all these $\psi$ states (with the exception of $\psi(3770)$) 
is that they are all above a $D \bar{D}$ threshold and yet this decay mode is
not a dominant one. This fact (among other things) raises the suspiction that 
these are not conventional $c \bar{c}$ states.

The nature of the $\psi(3770)$ resonance is still a subject of
debate. Conventionally, it has been regarded
as the lowest-mass D-wave charmonium state above the
$ D \bar{D}$ threshold, i.e. a pure $c \bar{c}$ meson in the quark model.
However, in Ref. \cite{vo05} it was suggested that the $\psi(3770)$ 
may contain a considerable tetraquark component. In that work it was also  
suggested that the tetraquark nature of the state would reveal itself in the 
decay $\psi(3770) \to \eta \, J/\psi$ and a prediction of the decay width in 
this channel was given. Very recently, this decay was observed by the BESIII 
collaboration \cite{bes3-23} and the measured width was close to the prediction 
made in \cite{vo05}, giving support to the possible tetraquark component of 
the $\psi(3770)$. In our formalism, we treat the vector mesons as $c \bar{c}$ 
bound states. So our predicted cross section refers to the production of a 
conventional charmonium or to the charmonium component of the mixed 
charmonium-tetraquark state.

The Particle Data Group (PDG) has been updating the parameters of vector
charmonium like states, $\psi(4160)$ and $ \psi(4230)$, thanks to
the improved data analysis techniques and the higher statistics of the data.  
The analysis done by BES reported higher Breit-Wigner (BW) mass for 
$\psi(4160)$:  $M = 4191.7 \pm 6.5$ MeV \cite{bes07}.
Although the updated mass and width parameters of these two states are closer 
to each other, they are commonly regarded as different states with the same 
quantum numbers whose underlying nature remains elusive. 
$\psi(4160)$ is regarded as a $2^3D_1$ $c \bar{c}$  state due to
its consistency with the predictions of the quark potential model \cite{qm}.
The $\psi(4160)$ and $\psi(4230)$ have the same quantum numbers with mass 
difference of about 40 MeV but can hardly be accommodated in
the quark model at the same time \cite{qm}. Furthermore, while the 
$\psi(4160)$ appears in the open charm channels, it is absent in the 
hidden-charm channels, and the
decay channels of $\psi(4230)$ listed in the PDG table are
mostly hidden-charm channels. Clearly, these states deserve further studies. 
In \cite{zlx23} it has been argued that 
the $\psi(4160)$ and $\psi(4230)$  are in fact the same state. 
The measurement of the production cross sections of these two states in 
the three photon fusion may help in elucidating their nature.

\begin{table}[h!]
\begin{center}
\begin{tabular}{|c|c|c|c|}
\hline
State & Mass & $\Gamma _{e^+ e^-}$[keV] &  $\sigma$[nb]
\tabularnewline
\hline
\hline
    $\rho ^0$ & 770          &  6.77 & 2466.9       \tabularnewline
    $\omega$  & 782          &  0.6  &  215.3        \tabularnewline
    $\text{J}/{\psi}$ & 3097 &  5.3  &  476.5    \tabularnewline
    $\psi(2S)$ & 3686        &  2.1  &  161.4  \tabularnewline
    $\psi$(3770) & 3770      &  0.26 &   19.5  \tabularnewline
    $\psi$(4040) & 4040      &  0.86 &   59.7  \tabularnewline
    $\psi$(4160) & 4160      &  0.48 &   32.4  \tabularnewline
    $\psi$(4230) & 4230      &  1.53 &  101.5 \tabularnewline
    $\psi$(4415) & 4415      &  0.58 &   36.9  \tabularnewline 
    \hline
\end{tabular}
\caption{Cross sections for production of C-odd  mesons in Pb-Pb
ultra-peripheral collisions at $\sqrt{s_{NN}}= 5.02$ TeV.
The decay widths are taken from the PDG \cite{pdg}.}
\label{tab4-23}
\end{center}
\end{table}

\section{Summary}

In this work we have studied the charmonium production in UPCs at LHC 
energies due to two and three photon fusion processes. 
These are clean processes where the particles of the 
initial state are intact at the final state and can be detected     
with the presence of two rapidity gaps between the                  
projectiles and the produced particle. We have used the QED formulas (derived 
in \cite{bn02}) complemented with the experimental data on decay widths.
We have predicted sizeable 
values for the cross sections in Pb-Pb collisions. We conclude that the  
experimental study is worth pursuing, that it can be useful to constrain 
decay widths evaluated theoretically and, ultimately, it can help 
in determining the structure of the considered states, confirming or not 
their quark-antiquark nature. 

\eject

\section{Appendix: equivalent photon approximation for two-photon processes}

In the equivalent photon approximation,
the cross section for the production of a generic  charmonium
state, $R$,    in UPCs between two hadrons, $h_{1}$ and $h_{2}$, is given
by (See e.g. \cite{BB88,hencken})
\begin{eqnarray}
\sigma \left( h_1 h_2 \rightarrow h_1 \otimes R \otimes h_2 ;s \right)
&=& \int \hat{\sigma}\left(\gamma \gamma \rightarrow R ;
W \right )  N\left(\omega_{1},{\mathbf b_{1}}  \right )
 N\left(\omega_{2},{\mathbf b_{2}}  \right ) S^2_{abs}({\mathbf b})
 \mbox{d}^{2} {\mathbf b_{1}}
\mbox{d}^{2} {\mathbf b_{2}}
\mbox{d} \omega_{1}
\mbox{d} \omega_{2} \,\,\, ,
\label{sec_hh}
\end{eqnarray} 
where $\sqrt{s}$ is center - of - mass energy for the $h_1 h_2$         
collision ($h_i = p, A$), $\otimes$ characterizes a rapidity gap in the   
final state and $W = \sqrt{4 \omega_1 \omega_2}$ is the invariant mass of 
the $\gamma \gamma$ system. The quantity $N(\omega_i,b_i)$ is the equivalent 
photon spectrum generated by hadron (nucleus) $i$, and 
$\sigma_{\gamma \gamma \rightarrow R}(\omega_{1},\omega_{2})$                
is the cross section for the production of a state $R$ from two real photons 
with energies $\omega_1$ and $\omega_2$. Moreover, in Eq. (\ref{sec_hh}), 
$\omega_{i}$ is the energy of the photon emitted by the hadron (nucleus) 
$h_{i}$ at an impact parameter, or distance, $b_{i}$ from $h_i$. Finally,   
in the above formula  $ S^2_{abs}({\mathbf b})$ is the survival probability 
written as the square of the scattering matrix,                      
introduced here to enforce the absorption at small impact parameters 
$ b \lesssim R_{h_1} + R_{h_2}$ \cite{BN93}. 
The equivalent photon flux can be expressed  as follows
\begin{equation}
N(\omega,b) = \frac{Z^{2}\alpha_{em}}{\pi^2}\frac{1}{b^{2}\omega}
\left[ \int u^{2} J_{1}(u) F\left(\sqrt{\frac{\left( {b\omega}/
{\gamma}\right)^{2} + u^{2}}{b^{2}}} \right )
\frac{1}{\left({b\omega}/{\gamma}\right)^{2} + u^{2}} \mbox{d}u\right]^{2}\, ,
\label{fluxo}
\end{equation}
where $F$ is the nuclear form factor of the  equivalent photon source.
In order to estimate the $ h_1 h_2 \rightarrow h_1 \otimes R \otimes h_2$ cross
section one needs the $\gamma \gamma \rightarrow R$ interaction cross section as 
input. Usually on uses the Low formula \cite{Low}, where the cross section
for the production of  the $R$ state due to the two-photon fusion can be written
in terms of the two-photon decay  width of  the corresponding state as
\begin{eqnarray}
 \sigma_{\gamma \gamma \rightarrow R}(\omega_{1},\omega_{2}) =
8\pi^{2} (2J+1) \frac{\Gamma_{R \rightarrow \gamma \gamma}}{M_{R}}
\delta(4\omega_{1}\omega_{2} - M_{R}^{2}) \, ,
\label{Low_cs}
\end{eqnarray}
where the decay width $\Gamma_{R \rightarrow \gamma \gamma}$ can in some cases 
be taken from experiment or can be theoretically estimated.                   
In the above formula, $M_{R}$ and $J$ are, respectively, the mass and spin of 
the produced state.

\begin{acknowledgments}
This work was  partially financed by the Brazilian funding agencies CNPq, FAPESP 
and also by the INCT-FNA and by the IANN-QCD network.
\end{acknowledgments}

\hspace{1.0cm}

\end{document}